\documentclass[aps,prc,twocolumn,floatfix]{revtex4}
\usepackage[dvips]{graphicx}
\usepackage{epsfig}
\usepackage{pst-plot}
\usepackage{bm}

\begin{document}
\title{Coupled cluster approach to nuclear physics}
\author{D.~J.~Dean}
\affiliation{Physics Division, Oak Ridge National Laboratory,
P.O. Box 2008, Oak Ridge, TN 37831-6373} 
\author{M.~Hjorth-Jensen}
\affiliation{Department of Physics and Center of Mathematics for Applications, University of Oslo, N-0316 Oslo, Norway}
\date{\today}
\begin{abstract}
Using many-body perturbation theory and 
coupled-cluster theory, we calculate the ground-state
energy of $^{4}$He and $^{16}$O. We perform these calculations 
using a no-core $G$-matrix interaction derived from a realistic 
nucleon-nucleon potential. Our calculations employ 
up to two-particle-two-hole coupled-cluster amplitudes.
\end{abstract}

\maketitle

\section{Introduction}
\label{sec:introduction}

The coupled-cluster method originated in nuclear physics over
forty years ago when Coester and Kummel proposed an exponential ansatz 
to describe correlations within a nucleus 
\cite{coester58,coester60}. This ansatz has been well justified
for many-body problems using a formalism in which the
cluster functions are constructed by cluster operators acting on 
a reference determinant \cite{harris92}. Early applications 
to finite nuclei were described in Ref.~\cite{kum78}. 
From that
time to this, a systematic development and implementation 
of this interesting many-body theory in nuclear 
physics applications has been only sporadic. The view from 
computational quantum chemistry is quite different. 
In fact, coupled-cluster methods applied to computational chemistry 
enjoy tremendous success 
\cite{bartlett81,comp_chem_rev00,piotr1,helgaker,arponen97,lindgren} 
over a broad class of
chemistry problems related to chemical and 
molecular structure and chemical reactions. 

Many solid theoretical reasons exist that motivate a pursuit of
coupled-cluster methods. First of all, the method is fully 
microscopic and is capable of systematic and hierarchical improvements.
Indeed, when one expands the cluster operator in coupled-cluster theory
to all $A$ particles in the system, one exactly produces the fully-correlated
many-body wave function of the system. The only input that the method
requires is the nucleon-nucleon interaction. 
The method may also be extended
to higher-order interactions such as the three-nucleon interaction. 
Second, the method is size extensive which means that only linked 
diagrams appear in the computation of the 
energy (the expectation value of the Hamiltonian) and amplitude equations. 
As discussed in Ref.~\cite{comp_chem_rev00} all shell model calculations 
that use particle-hole truncation schemes
actually suffer from the inclusion of unconnected diagrams 
in computations of the energy. 
Third, coupled-cluster theory is also size
consistent which means that the energy of two non-interacting fragments 
computed separately is the same as that computed for both fragments
simultaneously. In chemistry, where the study of reactions 
is quite important, this is a crucial property not available
in the interacting shell model (named configuration interaction in
chemistry).
Fourth, while the theory 
is not variational, 
the energy behaves as a variational quantity in most instances. 
Finally, from a 
computational point of view, the practical implementation of coupled 
cluster theory is amenable to parallel computing. 

Applications to nuclear problems resurfaced a few
years ago in the works of Mihalia and Heisenberg \cite{mh00a,mh00b,mh99,hm99}.
These efforts focused primarily on the structure of
$^{16}$O, and used a strategy of solution that is somewhat different
from the approach we will take in this and subsequent articles. One 
major difference is that we will use a $G$-matrix to renormalize the
two-body interactions before we begin our coupled-cluster calculations. 
We will also take a somewhat different approach in our Hilbert 
space truncation. Also notable is the work of Moliner, Walet, and Bishop
\cite{mwb02} who are pursuing nuclear problems in translationally 
invariant coupled cluster methods in coordinate space.

The computed energy using the coupled-cluster formalism includes 
a very large class of many-body perturbation theory diagrams.
In standard many-body perturbation theory, one typically sums 
all diagrams order by order. The coupled-cluster approach essentially
iterates diagrams so that one may discuss it in terms of an infinite 
summation of particular classes of diagrams.  Thus the theory is
nonperturbative. In fact the 
coupled-cluster energy at the single and double excitation level 
contains contributions identical to 
those of second order and third order many-body perturbation theory, but
lacks triple excitation contributions necessary to complete fourth-order
many-body perturbation theory; see e.g., the review article of 
Bartlett \cite{bartlett81}. It has been shown that the quadruple
excitation contributions may be factored exactly
into products of double excitations, but no such factorization is 
possible for the corresponding triples. Therefore, the coupled-cluster energy 
lacks only triple excitation contributions to be complete through 
fourth order.  

In this paper, we wish to establish a line of research that we
intend to pursue for calculating nuclear properties using coupled-cluster 
techniques. This is therefore a first paper in a series that 
we will publish to both develop the method for nuclear physics and to 
demonstrate the power of the method for various applications. This 
first installment will be devoted to outlining our approach, investigating
the physical motivations, establishing numerical convergence tests, and 
presenting some initial calculations using the method. 
In Sec.~\ref{sec:gmatrix},
we will describe our choice of reduced Hilbert space, 
construction of an effective interaction for various model spaces 
and elimination of the spurious center of mass motion. The model spaces
are defined in terms of various major harmonic oscillator shells and the 
effective interaction is defined in terms of the $G$-matrix,
see e.g., Ref.~\cite{hko95}. 
Our calculations of the binding energy of nuclei like helium and oxygen
entail therefore a dependence upon the number of harmonic oscillator states,
the oscillator parameter 
and the starting energy at which the $G$-matrix is computed. 
The convergence of the binding energy as function of the number of harmonic
oscillator shells  is a crucial test of the method discussed in this work. 
Since the coupled-cluster calculations are rather time consuming for systems
like $^{16}$O, we present, as an introduction to the coupled-cluster method, 
intermediate results from 
perturbative many-body approaches in Sec.~\ref{sec:mbpt}. 
Sec.~\ref{sec:mbpt} also
serves the purpose of finding an eventual minimum for the energy 
as function of the oscillator parameter, with which we limit the 
number of coupled-cluster computations. It may also be of help  
in finding  how many oscillator shells
are needed in order to achieve a satisfactory convergence. In addition, 
this section sheds lights on the limitations of many-body perturbation theory compared 
with the coupled-cluster approach.

We turn in Sec.~\ref{sec:ccsd}  to a 
description of the coupled-cluster equations. 
We discuss the numerical techniques
we will employ to obtain solutions to the coupled-cluster
equations and demonstrate several results 
for $^{4}$He and $^{16}$O in Sec.~\ref{sec:results}.
We conclude with a prospective for future directions of this research
in Sec.~\ref{sec:conclusion}. 

\section{Effective interactions for a truncated Hilbert space}
\label{sec:gmatrix}
The aim of this section is to present and partly justify the computation of 
an effective two-body Hamiltonian acting within a reduced Hilbert space.
This two-body Hamiltonian will in turn serve as the starting point for the
perturbative approach of Sec.~\ref{sec:mbpt} and the 
coupled-cluster expansion discussed in Sec.~\ref{sec:ccsd}.

Before we can compute such an effective two-body Hamiltonian,
we need to define the nucleon-nucleon interaction.    
Several types of modern nucleon-nucleon scattering interactions
have been developed during recent years. These interactions all 
fit nucleon-nucleon scattering data up to 300 MeV with excellent
precision \cite{cdbonn,bob95,stoks94}. They do give slightly differing
results for the radius of the deuteron, the binding energy of the triton
and also contain slight differences in the way they treat locality. 

Very recent work by Entem and Machleidt \cite{machleidt02}
provides for the first time an interaction of quantitative accuracy
that is based on effective field theory. One basic open question of 
nuclear theory involves understanding how the nucleon-nucleon interaction
may be derived from quantum chromodymanics, 
the theory of strong interactions. Quantum chromodymanics has
not been solved in its nonperturbative low-energy limit at energy scales
that are characteristic for low-energy nuclear physics. One promising
way to circumvent this problem is to employ a derivation of the nuclear
force based on chiral effective field theory 
\cite{weinberg,vankolck}. The authors of Ref.~\cite{machleidt02} 
undertook the task of generating an accurate
nucleon-nucleon interaction  based on chiral perturbation theory. They
included one- and two-pion exchange contributions up to 
chiral order three. They also showed that a quantitative fit of the 
nucleon-nucleon $D$-wave phase shifts requires contact terms representing
short-range forces of order four. The number of free parameters used in
this chiral interaction is 46, which is similar to the number of free
parameters found in other two-nucleon forces. The phase-shift analysis
shows excellent agreement between the chiral interaction and the scattering
data. 

Two interactions were formulated in Ref.~\cite{machleidt02}. These two 
models, denoted Idaho-A and Idaho-B, differ in their prediction of the
$D$-state probabilities of the deuteron, while both interactions will 
give the same values for the $^3S_1$, $^3D_1$ and $\epsilon_1$ phase 
parameters up to 300~MeV in scattering energy. Idaho-A yields a $D$-state
probability of 4.17\%, while Idaho-B gives 4.94\%. This also affects
the triton binding energy, yielding 8.14~MeV and 8.02~MeV for Idaho-A
and Idaho-B, respectively. 
A similar interaction which now goes to  
fourth order in chiral perturbation theory and includes charge dependence,  
has recently been presented by Entem and Machleidt, see Ref.~\cite{machleidt03}.
Since this is a methodological paper, we limit the attention to one of 
these interactions, namely the Idaho-A model. Results for other interactions, such as
the $V_{18}$ model of the Argonne group \cite{bob95}, 
will be presented in future work.

\subsection{Definition of the model space and the two-body effective interaction}

In order to derive an effective interaction suitable for coupled
cluster calculations, we need to introduce various
notations and definitions pertinent to the methods exposed.

A common practice in nuclear many-body theory is to reduce the infinitely
many degrees of freedom of the Hilbert space to those represented
by a physically motivated subspace, the model space.
In such truncations of the Hilbert space, the notions of a projection
operator $P$ onto the model space and its complement $Q$ are
introduced. The projection operators defining the model and excluded
spaces are
\begin{equation}
        P=\sum_{i=1}^{D} \left|\Phi_i\right\rangle
        \left\langle\Phi_i\right |,
\label{eq:poperator}
\end{equation}
and
\begin{equation}
        Q=\sum_{i=D+1}^{\infty} \left|\Phi_i\right\rangle
        \left\langle\Phi_i\right |,
\label{eq:qoperator}
\end{equation}
with $D$ being the dimension of the model space, and $PQ=0$, $P^2 =P$,
$Q^2 =Q$ and $P+Q=I$. The two-body wave functions $\left|\Phi_i\right\rangle$ 
are normally eigenfunctions
of an unperturbed Hamiltonian $H_0$. In this work we let only the kinetic energy
enter the definition of $H_0$, i.e., $H_0=t$. 
Since we will employ a harmonic oscillator basis, this means that 
we need to  compute  the expectation value of $H_0$ as well. The unperturbed wave functions are not eigenfunctions of $t$. 
The full Hamiltonian
is then $H=t+V_{NN}$ with $V_{NN}$ the
nucleon-nucleon interaction.
The eigenfunctions of the full two-body Hamiltonian are denoted by
$\left|\Psi_{\alpha}\right\rangle$
and $E_{\alpha}$,
\begin{equation}
                H\left|\Psi_{\alpha}\right\rangle= 
                E_{\alpha}\left|\Psi_{\alpha}\right\rangle.
\end{equation}
Rather than solving the full Schr\"{o}dinger equation above, we define
an effective Hamiltonian acting within the model space such
that
\begin{equation}
               PH_{\mathrm{eff}}P\left|\Psi_{\alpha}\right\rangle=
               E_{\alpha}P\left|\Psi_{\alpha}\right\rangle=
              E_{\alpha}\left|\Phi_{\alpha}\right\rangle
\end{equation}
where $\left|\Phi_{\alpha}\right\rangle=P\left|\Psi_{\alpha}\right\rangle$
is the projection of the full wave function
onto the model space, the model space wave function.
Here $H_{\mathrm{eff}}$  is an effective Hamiltonian acting solely
within the chosen model space
given by $H_{\mathrm{eff}}=PtP+V_{\mathrm{eff}}$, with the interaction
\begin{equation}
  V_{\mathrm{eff}}=\sum_{i=1}^{\infty} V_{\mathrm{eff}}^{(i)},
\end{equation}
where $ V_{\mathrm{eff}}^{(1)}$,  $ V_{\mathrm{eff}}^{(2)}$,
 $ V_{\mathrm{eff}}^{(3)}$,...\ are effective one-body, two-body,
three-body interactions etc. For finite $A$-body systems, the sum terminates
at $i=A$. As stated above, in this work we will limit the
attention to two-body interactions.
The next step could be to employ perturbative many-body techniques 
or the coupled 
cluster method.
In  perturbation theory, the effective interaction
$H_{\mathrm{eff}}$ can be written out order by order in the 
interaction $V_{NN}$ as
\begin{widetext}
\begin{equation}
               PH_{\mathrm{eff}}P=PtP+PV_{NN}P +PV_{NN}\frac{Q}{e}V_{NN} P+
               PV_{NN}\frac{Q}{e}V_{NN} \frac{Q}{e}V_{NN} P+\dots.
               \label{eq:effint}
\end{equation}
\end{widetext}
In this expansion, $e=\omega -t$,
where $\omega$ is the so-called starting energy, defined as the unperturbed
energy of the interacting particles.
However, 
the nucleon-nucleon interactions all possess a hard core that 
makes them unsuitable for perturbative many-body approaches. The
standard procedure is therefore to  renormalize the short-range part of the 
interaction by introducing the so-called reaction matrix $G$
\begin{equation}
G=V_{NN}+V_{NN}\frac{\tilde{Q}}{\omega - \tilde{Q}t\tilde{Q}}G.
\label{eq:gfinite}
\end{equation}
The operator $\tilde{Q}$ is normally different from the 
projection operator defined in Eq.~(\ref{eq:qoperator}), 
since the $G$-matrix by construction 
allows only specific two-body states to be defined by $\tilde{Q}$.
Typically,  
the $G$-matrix is the sum over all
ladder type of diagrams with intermediate particle-particle states
only. This sum is meant to renormalize
the repulsive short-range part of the interaction. The physical interpretation
is that the particles must interact with each other an infinite number
of times in order to produce a finite interaction. This interaction can in turn
serve as an effective interaction acting in a reduced space. 

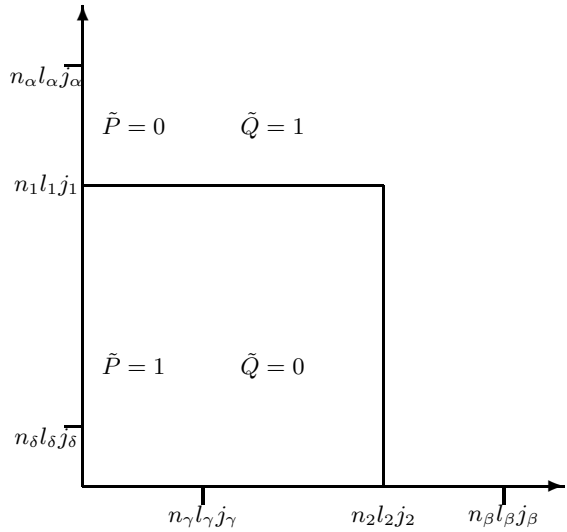
\begin{figure}[htbp]
\begin{center}
\setlength{\unitlength}{0.8cm}
\begin{picture}(9,10)
\thicklines
   \put(1,0.5){\makebox(0,0)[bl]{
              \put(0,1){\vector(1,0){8}}
              \put(0,1){\vector(0,1){8}}
              \put(7,0.5){\makebox(0,0){$n_{\beta}l_{\beta}j_{\beta}$}}
              \put(7,0.7){\line(0,1){0.3}}
              \put(2,0.5){\makebox(0,0){$n_{\gamma}l_{\gamma}j_{\gamma}$}}
              \put(2,0.7){\line(0,1){0.3}}
              \put(-0.6,6){\makebox(0,0){$n_1l_1j_1$}}
              \put(5,0.5){\makebox(0,0){$n_2l_2j_2$}}
              \put(2,3){\makebox(0,0){$\tilde{P}=1\hspace{1cm}\tilde{Q}=0$}}
              \put(2,7){\makebox(0,0){$\tilde{P}=0\hspace{1cm}\tilde{Q}=1$}}
              \put(-0.6,7.8){\makebox(0,0){$n_{\alpha}l_{\alpha}j_{\alpha}$}}
              \put(-0.3,8){\line(1,0){0.3}}
              \put(-0.6,1.8){\makebox(0,0){$n_{\delta}l_{\delta}j_{\delta}$}}
              \put(-0.3,2){\line(1,0){0.3}}
              \put(0,6){\line(1,0){5}}
              \put(5,1){\line(0,1){5}}
         }}
\end{picture}
\caption{Definition of the exclusion operator used to compute the $G$-matrix. 
See text for further details. \label{fig:paulioperator}}
\end{center}
\end{figure}

We illustrate the definition of the exclusion operator employed in this
work in Fig.~\ref{fig:paulioperator}. 
Using a harmonic oscillator basis for the single-particle wave functions,
a single-particle state is classified by the quantum numbers $nlj$.
A two-particle state in an angular momentum coupling scheme is given by
$\left| (n_{\alpha}l_{\alpha}j_{\alpha}n_{\beta}l_{\beta}j_{\beta})JT_Z\right \rangle$, where $\alpha$ and $\beta$ represent one of the orbitals 
$0s_{1/2}$, $0p_{3/2}$, $0p_{1/2}$ etc and $J$ is the total two-particle
angular momentum and $T_Z$ the corresponding isospin projection.

The single-particle states labeled by $n_1l_1j_1$ and $n_2l_2j_2$
represent the last orbit of model space $\tilde{P}$. 
In this work  $n_1l_1j_1$ and $n_2l_2j_2$ will mark the number of harmonic
oscillator shells included in the definition of $\tilde{P}$. In the actual
calculations presented below these range from four to eight major shells.
For four major shells $n_1l_1j_1=2s_{1/2}$ and $n_2l_2j_2=2s_{1/2}$
while for eight major shells we get
$n_1l_1j_1=3p_{1/2}$ and $n_2l_2j_2=3p_{1/2}$ as the last single-particle
orbits.
In Fig.~\ref{fig:paulioperator} the two-body state
$\left| (n_{\alpha}l_{\alpha}j_{\alpha}n_{\beta}l_{\beta}j_{\beta})JT_Z\right \rangle$ does not belong to the model space and is included 
in the computation of
the $G$-matrix. 
Similarly, 
$\left| (n_{\alpha}l_{\alpha}j_{\alpha}n_{\gamma}l_{\gamma}j_{\gamma})JT_Z\right \rangle$
and 
$\left| (n_{\delta}l_{\delta}j_{\delta}n_{\beta}l_{\beta}j_{\beta})JT_Z\right \rangle$ 
also enter the definition of $\tilde{Q}$ whereas 
$\left| (n_{\delta}l_{\delta}j_{\delta}n_{\gamma}l_{\gamma}j_{\gamma})JT_Z\right \rangle$ 
is not included in the computation of $G$. 
This means that correlations not defined in the $G$-matrix need 
to be computed by other non-perturbative
resummations or many-body schemes. 
This is where the coupled-cluster scheme enters.

Before we proceed we outline the computation of the $G$-matrix using the 
exclusion operator of Fig.~\ref{fig:paulioperator}.
One can solve the equation for the $G$-matrix
for finite nuclei by employing
a formally
exact technique for handling $\tilde{Q}$
discussed in e.g., Ref.~\cite{hko95}.
Using the matrix identity, for which $\tilde{P}$ is the complement of 
$\tilde{Q}$ such that $\tilde{P}+\tilde{Q}=1$,
\begin{equation}
  \tilde{Q}\frac{1}{\tilde{Q}e\tilde{Q}}
  \tilde{Q}=\frac{1}{e}-
   \frac{1}{e}\tilde{P}\frac{1}{\tilde{P}e^{-1}\tilde{P}}\tilde{P}\frac{1}{e},
   \label{eq:matrix_relation_q}
\end{equation}
to rewrite Eq.~(\ref{eq:gfinite}) 
as
\begin{equation}
   G = G_{F} +\Delta G,\label{eq:gmod}\;,
\end{equation}
where $G_{F}$ is the free $G$-matrix defined as
\begin{equation}
   G_{F}=V_{NN}+V_{NN}\frac{1}{\omega - t}G_{F}. \label{eq:freeg}
\end{equation}
The term $\Delta G$ is a correction term defined entirely within the
model space $\tilde{P}$ and given by
\begin{equation}
   \Delta G =-V_{NN}\frac{1}{A}\tilde{P}
   \frac{1}{\tilde{P}A^{-1}\tilde{P}}\tilde{P}\frac{1}{A}V_{NN}.
\end{equation}
Employing the definition for the free $G$-matrix of Eq.\ (\ref{eq:freeg}),
one can rewrite the latter equation as
\begin{equation}
  \Delta G =-G_{F}\frac{1}{e}\tilde{P}
  \frac{1}{\tilde{P}(e^{-1}+e^{-1}G_{F}e^{-1})
  \tilde{P}}\tilde{P}\frac{1}{e}G_F,
\end{equation}
with $e=\omega -t$.
We see then that the $G$-matrix 
is expressed as the sum of two
terms; the first term is the free $G$-matrix with no  corrections
included, while the second term accounts for medium modifications
due to the exclusion operator $\tilde{Q}$. The second term can easily
be obtained by some simple matrix operations involving
the model-space matrix $\tilde{P}$ only. The above allows, for a given 
model space operator
$\tilde{P}$, for a numerically  exact computation of the $G$-matrix. 

The $G$-matrix defined by the exclusion operator $\tilde{Q}$
of Fig.~\ref{fig:paulioperator} represents now our effective two-body
interaction, with a reduced Hilbert space defined by the model space
$\tilde{P}$.  This means that our Hamiltonian acts within the model 
space $\tilde{P}$ and is given by 
\begin{equation}
H_{\mathrm{eff}}(\omega) = t+ G(\omega).
\label{eq:finalheff}
\end{equation}
The results from exact shell-model diagonalizations, perturbative
many-body calculations and coupled-cluster calculations will depend
on the limits  of the model space $\tilde{P}$ 
and the chosen oscillator parameter.  Furthermore, although the $G$-matrix has a 
weak dependence upon the starting energy $\omega$, this dependence will affect
the final results the calculations. 
The dependence of the results
upon these  parameters will be elucidated in Secs.~\ref{sec:mbpt} and 
\ref{sec:results}. 

We end this subsection with the setup of the exclusion operator used in the final
perturbative many-body calculations of Sec.~\ref{sec:mbpt} and the coupled
cluster  calculations of Sec.~\ref{sec:results}.   

\begin{figure}[htbp]
\begin{center}
\setlength{\unitlength}{0.8cm}
\begin{picture}(9,10)
\thicklines
   \put(1,0.5){\makebox(0,0)[bl]{
              \put(0,1){\vector(1,0){8}}
              \put(0,1){\vector(0,1){8}}
              \put(7,0.5){\makebox(0,0){b}}
              \put(-0.6,6){\makebox(0,0){$n_pl_pj_p$}}
              \put(5,0.5){\makebox(0,0){$n_ql_qj_q$}}
              \put(-0.6,3){\makebox(0,0){$n_il_ij_i$}}
              \put(2,0.5){\makebox(0,0){$n_il_ij_i$}}
              \put(-0.6,8){\makebox(0,0){a}}
              \put(0,6){\line(1,0){5}}
              \put(5,1){\line(0,1){5}}
              \put(0,3){\line(1,0){2}}
              \put(2,1){\line(0,1){2}}
         }}
\end{picture}
\caption{Definition of particle and hole states 
for coupled-cluster and perturbative many-body calculations. 
We use the notation that
$p,q,r,s$ index all single-particle states, while $i,j,k,l$ refer to single-hole states
The orbits represented by the quantum numbers 
 $n_il_ii_i$ represent hole states whereas $n_{p,q}l_{p,q}j_{p,q}$ 
represent the last particle orbits included in the $G$-matrix model space. 
The hole states define the 
Fermi energy.\label{fig:finalp}}
\end{center}
\end{figure}
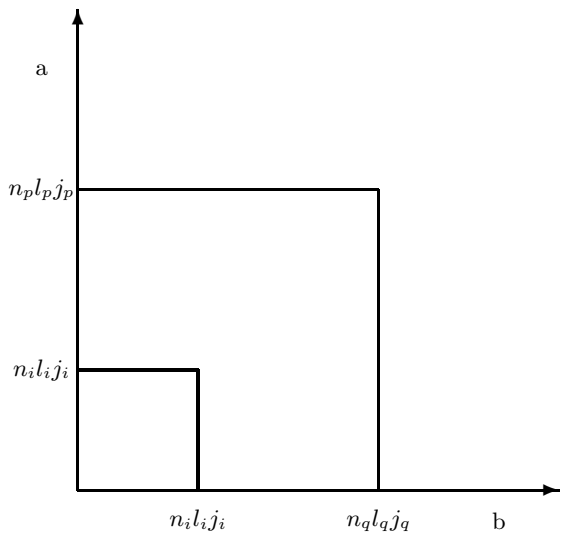

With the $G$-matrix model space $\tilde{P}$ of Fig.~\ref{fig:paulioperator} we can now define
an appropriate space for many-body perturbation theory 
and coupled-cluster calculations where correlations
not included in the $G$-matrix are to be generated. This model space is defined 
in Fig.~\ref{fig:finalp}, where the labels $n_{p,q}l_{p,q}j_{p,q}$ represent the same
single-particle orbits as $n_{1,2}l_{1,2}j_{1,2}$ in  Fig.~\ref{fig:paulioperator}. 
Hereafter we use the notation that
$p,q,r,s$ index all single-particle states, while $i,j,k,l$ refer to single-hole states

The $G$-matrix does not reflect a specific nucleus and 
thereby single-particle orbits which define the uncorrelated 
Slater determinant.  For a nucleus like 
$^{4}$He the $0s_{1/2}$ orbit is fully occupied and defines thereby single-hole states.
These are labeled by $n_il_ij_i$ in Fig.~\ref{fig:finalp}. 
For $^{16}$O the corresponding hole states are represented by the orbits
$0s_{1/2}$,  $0p_{3/2}$ and  $0p_{1/2}$. With this caveat we can then generate
correlations not included in the $G$-matrix. 

The unperturbed Hamiltonian $H_0$ is given by  the kinetic energy only.
However, in order to define a suitable starting point for 
an effective interaction to be used
in the coupled-cluster calculations we use the 
definition of particles and holes in 
Fig.~\ref{fig:finalp} and compute the single-hole energies $\varepsilon_i$ through 
\begin{equation}    \label{eq:poormansbhf}
  \varepsilon_i = \left\langle i\right | \frac{p^2}{2m} \left | i \right \rangle +\sum_{j \leq F}  \left\langle ij\right | G(\omega=\varepsilon_i+\varepsilon_j) \left | ij\right\rangle, 
\end{equation}
where $F$ stands for the Fermi energy. We do not perform a self-consistent Brueckner-Hartree-Fock 
calculation however, as done by e.g., Gad and M\"uther \cite{herbert02}, the reason being that
this self-consistency is taken care of by the coupled-cluster procedure.
The matrix elements are all
antisymmetrized.
Furthermore, for single-particle states above the Fermi energy we leave the single-particle energies
unchanged. This procedure, which follows the Bethe-Brandow-Petschek theorem \cite{bbp63}, 
introduces an artificial gap at the Fermi surface. 
Note also that the single-particle wave functions are not changed in Eq.~(\ref{eq:poormansbhf}). 
The main purpose of the above procedure is to yield a 
prescription for obtaining
a starting energy independent effective interaction 
for the coupled-cluster 
calculations. Using the single-particle energies from 
Eq.~(\ref{eq:poormansbhf}) 
we define, following Ref.~\cite{herbert02}, an effective
interaction for our coupled-cluster model spaces by 
\begin{eqnarray}
    \left\langle ij \right | {\cal V}_{\mathrm{eff}} \left |
kl \right\rangle
&=& \frac{1}{2} \left[ \left\langle ij \right |
G(\omega=\varepsilon_i+\varepsilon_j) \left | kl \right\rangle \right. 
\nonumber \\ 
&+&  \left.\left\langle ij\right | 
G(\omega=\varepsilon_k+\varepsilon_l) \left |  kl\right\rangle \right], 
\label{eq:gmat1approx}
\end{eqnarray}
for two-body states with holes only and 
\begin{eqnarray}
    \left\langle ip \right | {\cal V}_{\mathrm{eff}}\left | kq \right\rangle 
&=& \frac{1}{2}\left[ \left\langle ip \right | G(\omega=\varepsilon_i
+\varepsilon_p) \left | kq \right\rangle \right. \nonumber \\
&+& \left. \left\langle ip \right | G(\omega=\varepsilon_k+\varepsilon_q) 
\left | kq \right\rangle \right], 
\label{eq:gmat2approx}
\end{eqnarray}
for two-body states with one particle and one hole. For two-body states with two single-particle
states $pq$ we use a fixed starting energy, typically in the range $\omega \in [-80,-5]$~MeV. 
This introduces a starting energy 
dependence in our results. The reason for fixing the starting
energies for two-particle states is due to the fact that we use kinetic energies only above the
Fermi surface and our $G$-matrices are computed at negative starting energies only.

An obvious improvement to this procedure is to generate a complex $G$-matrix which takes care
of both positive and negative energies, reflecting thereby the non-resonant continuum, bound two-body
states and eventual resonances and/or virtual states. We will however leave this interesting topic
to a further study. The main focus of this paper is to establish a method for doing
the coupled-cluster calculations without having to resort to employing the bare interaction,
needing thereby  many major harmonic oscillator shells. Our hope is that with a $G$-matrix, 
$7-8$ major shells may suffice. 
Such a truncation in the harmonic oscillator space is supported by the recent works of
Barrett, Navratil, and Vary \cite{bruce1,bruce2,bruce3}, 
see also the recent calculations
of Ref.~\cite{petr_erich02}. In these works no-core shell-model calculations have 
been mounted
for light nuclei ranging from the triton to mass $A=12$. 
Furthermore, in Refs.~\cite{bruce1,bruce2},
another approach for obtaining a starting energy independent interaction is obtained through
the similarity transformations of Lee and Suzuki \cite{ls0,ls1}, 
yielding a fully hermitian
effective interaction. This should be contrasted to the more 
approximative method presented 
in Eqs.~(\ref{eq:gmat1approx}) and (\ref{eq:gmat2approx}).

\subsection{Treatment of center-of-mass motion}

Momentum conservation requires 
that a many-body wave function must factorize
as $\Psi(\bf{r}) = \phi(R)\Psi(\bf{r}_{\rm rel})$ where 
$R$ is the center-of-mass coordinate and $\bf{r}_{\rm rel}$ the 
relative coordinates. If we choose to expand our wave functions in
the harmonic oscillator basis, then we are able to exactly separate the
center-of-mass motion from the problem provided that we work in a model
space that includes all $n\hbar\omega$ excitations. 
In this paper, we perform calculations with a $Q$ operator that allows 
for all possible two-particle interactions within a given 
set of oscillator shells.  This means that we are 
$n\hbar\omega$ incomplete in a given 
calculation so that our method of separation of the center-of-mass 
motion becomes approximate. For example, for $^4$He in four major oscillator
shells, we can excite all particles to $n=12\hbar\omega$ excitations, 
but we can only excite one particle to $n=3\hbar\omega$ excitations. 
Thus, care must be taken when correcting for center-of-mass contamination
in our calculations. We have taken a variational approach based on the 
the work of Whitehead {\em et al.}, \cite{whitehead}. The idea is 
to add $\beta_{\rm c.m.}H_{\rm c.m}$ to the Hamiltonian, but 
with $\beta_{\rm c.m.}$ remaining fairly
small. This minimizes the effects of the center-of-mass contamination on
low-lying state properties, and partially pushes unwanted states out of the
spectrum. If we were to use a large $\beta_{\rm c.m.}$, we would find 
spurious states entering into the calculated low-lying spectrum due to 
the incompleteness of our model space. 

We proceed as follows. 
The center-of-mass Hamiltonian is 
\begin{equation}
H_{\rm c.m} = \frac{{\bf P}^2}{2MA} 
+\frac{1}{2}mA\omega^2{\bf R}^2 - \frac{3}{2}\hbar\omega \;,
\end{equation}
where ${\bf P}=\sum_{i=1,A}{\bf p}_i$ and $R=(\sum_{i=1,A}{\bf r}_i)/A$. 
$H_{\rm c.m.}$ can be rewritten as a one-body harmonic potential, and a
two-body term that depends on both the relative and center-of-mass 
coordinates of the two interacting particles. The matrix elements for
the two body terms may be found in Ref.~\cite{lawson}. Operationally, 
we add $H_{\rm c.m}$ to our Hamiltonian 
\begin{equation}
H' = H+\beta_{\rm c.m.}H_{\rm c.m}\;,
\end{equation}
where we choose $\beta_{\rm c.m}$ so that the expectation value of 
$H_{\rm c.m}$ is zero \cite{dean99}. This 
insures that our center-of-mass contamination 
within the many-body wave function is minimized. We also find that this
procedure yields reasonable spectra (in a space of four major
oscillator shells) for ${^4}$He \cite{papenbrock03}. 

\section{Perturbative many-body methods}
\label{sec:mbpt}
 Our results will depend on the size of the model space and the chosen harmonic
oscillator energy $\hbar\omega$.
This section serves therefore two aims central to the coupled-cluster approach of this work.
\begin{itemize}
\item In the coupled-cluster calculations we search for an energy
minimum as function of e.g., the oscillator energy $\hbar\omega$. Even for small systems 
like  $^{16}$O with five or six 
major shells included, these are major time-consuming calculations.
Results from many-body perturbation theory, which to second or third order 
in the interaction $G$ are fairly
simple, may therefore serve as a guide in order to limit the range of $\hbar\omega$ values
used in the coupled-cluster calculations.
With increasing dimensionality of the $G$-matrix model space, the results should become
independent of the oscillator energy.
\item  The energy will also depend on the size of the model space. The hope is that not too
many shells are needed in order to achieve a converged energy. 
With the  present approach, our 
coupled-cluster calculations of  
$^{16}$O are limited to six major shells.
Results from many-body perturbation theory can thus serve as a guideline.
\end{itemize}

\begin{figure}[hbpt]
\vspace{0.25in}
\includegraphics[scale=0.35]{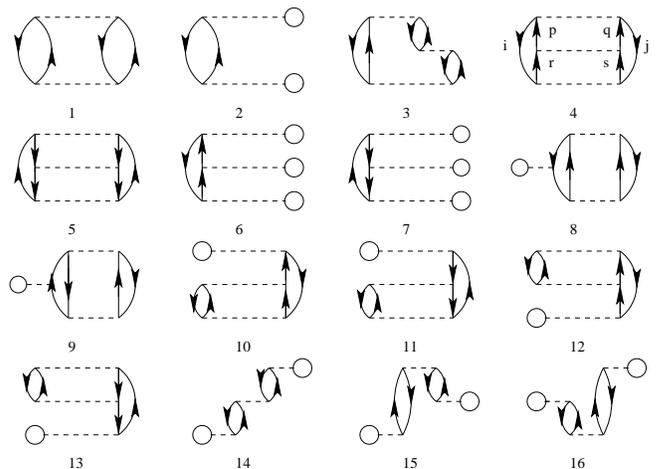}
\caption{Antisymmetrized Goldstone diagrams through third order in perturbation 
theory included
in the evaluation of the binding energy. The dashed lines represents the interaction, in our case the 
$G$-matrix.
Particle and hole states are represented by upward  and downward arrows, respectively.
The first order diagram is omitted. All closed circles stand for a summation
over hole states.}
\label{fig:diagrams}
\end{figure}

The linked-diagram theorem \cite{blaizot,lindgren} can be used to obtain a perturbative 
expansion  for the energy
in terms of the perturbation $V(G)$ or  $V=H-H_0$ where $H_0$ represents the unperturbed 
part of the Hamiltonian. The expression for the energy $E$ reads
\begin{equation}
  E = \sum_{k=0}^{\infty} \left\langle \Psi_0 \right| H\left[(\omega - H_0)^{-1}H\right]^k
      \left|\Psi_0\right\rangle_{L},
\end{equation}
where $\Psi_0$ is the uncorrelated Slater determinant for the ground state, $\omega$ is 
the corresponding  
unperturbed  energy and the subscript $L$ stands for linked diagrams only.
In our calculations, we must replace the Hamiltonian $H$ with the effective one defined
in Eq.~(\ref{eq:finalheff}) and employ the definition of particle and hole states
of Fig.~\ref{fig:finalp}.
In Fig.~\ref{fig:diagrams} we show all antisymmetrized Goldstone diagrams through third order
in perturbation theory (we omit the first order diagram). All closed circles stand for a summation
over hole states. In this section we let the  $G$-matrix define the interaction 
vertex. Since we do not use a self-consistently determined single-particle Hamiltonian, we
need to account for diagrams with so-called 
Hartree-Fock insertions as well. Examples of the latter are
shown in Fig.~\ref{fig:diagrams}, see also the work of Kassis \cite{kassis72} for a detailed
discussion of the various diagrams.
There is an additional problem with our 
many-body perturbation theory calculations. Consider the expression $\Delta E_4$
for diagram 4 in an angular momentum coupled basis (with  $J$ being the total two-body angular
momentum and $T_z$ the corresponding isospin projection)
of Fig.~\ref{fig:diagrams}
\begin{widetext}
\begin{eqnarray}
\Delta E_4 &=& \frac{1}{8}\sum_{\begin{array}{c}ij\leq F\\ pqrs > F\\ J\end{array}}(2J+1)
\left\langle (ij)JT_z\right | G(\omega=\varepsilon_i+\varepsilon_j) \left | (pq)JT_z 
\right\rangle\frac{1}{\varepsilon_i+\varepsilon_j-\varepsilon_p-\varepsilon_q} \\ \nonumber 
&& \times \left\langle (pq)JT_z\right | G(\omega=\varepsilon_i+\varepsilon_j) \left | (rs)JT_z
\right\rangle\frac{1}{\varepsilon_i+\varepsilon_j-\varepsilon_r-\varepsilon_s}
\left\langle (rs)JT_z\right | G(\omega=\varepsilon_i+\varepsilon_j) \left | (ij)JT_z
\right\rangle.
\end{eqnarray}
\end{widetext}
The $G$-matrices depend on the starting energy $\omega$, defined 
in this case by the hole energies.
With a harmonic oscillator basis and with kinetic energies only 
these starting energies will be positive, whereas
our $G$-matrix is defined for negative energies only with  $\omega \in [-140,-5]$ MeV.
To remove this problem we employ the  single-hole energies computed
according to Eq.~(\ref{eq:poormansbhf}). This leads to an artificial gap at the Fermi surface
and demonstrates one of the problems with many-body perturbation theory. A self-consistent 
approach is needed both for holes and particles, and the $G$-matrix needs to be 
computed for both
positive and negative energies. This is however beyond the scope of this work, where our
emphasis is on the coupled-cluster approach.  The center-of-mass corrections discussed in the previous
section are not included in our perturbative calculations.
This section, as will be seen below, serves the main aim
of justifying our model space used in the coupled-cluster computation. 
In addition, the energy minimum as function of the oscillator energy limits the range
of $\hbar\omega$ values used in the coupled-cluster calculations. 
 
Needless to say, many-body perturbation theory 
has severe limitations. It is very difficult to go beyond third order
in perturbation theory 
without a self-consistent single-particle potential and beyond 
fourth order with a self-consistent potential.  
The coupled-cluster method offers on the other hand a systematic way to generate all
many-body correlations in a given model space. 

\subsection{Many-body perturbation theory results for helium and oxygen}
\begin{figure}[hbtp]
\vspace{0.25in}
\includegraphics[angle=270, scale=0.35]{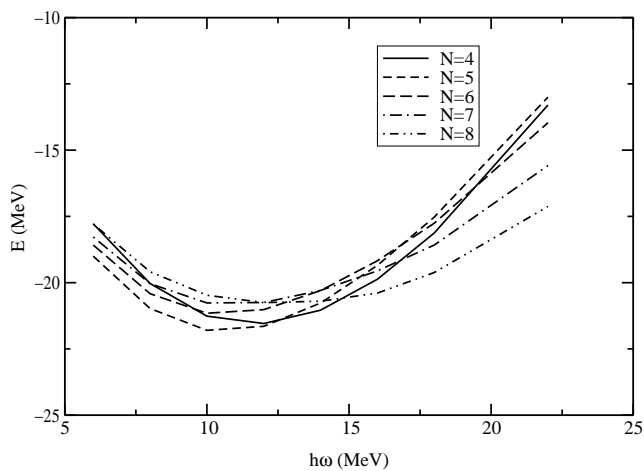}
\caption{Binding energy $E$ 
from third-order perturbation theory for $^{4}$He as function
of the number of major harmonic oscillator shells $N$ 
and the oscillator energy $\hbar\omega$. For $N=8$ we have the optimal value of
$E=-20.830$ MeV at $\hbar\omega = 13.3$ MeV. The experimental value is
$E=-28$ MeV.}
\label{fig:mbpthe}
\end{figure}

\begin{figure}
\vspace{0.25in}
\includegraphics[angle=270, scale=0.35]{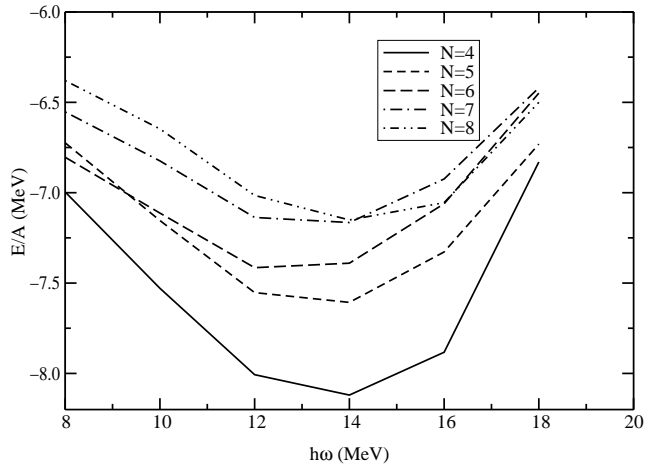}
\caption{Binding energy per particle $E/A$ 
from third-order perturbation theory for $^{16}$O as function 
of the number of major harmonic oscillator shells $N$ 
and the oscillator energy $\hbar\omega$. For $N=8$ we have the optimal value of
$E/A=-7.120$ MeV at $\hbar\omega = 13.6$ MeV. The experimental value is
$E/A=-7.98$ MeV.}
\label{fig:mbptox}
\end{figure}

We present here results from third-order in perturbation theory for the 
binding energies of $^{4}$He and $^{16}$O as functions of the size of the model
space and the chosen oscillator energy $\hbar\omega$.
These results are shown in Figs.~\ref{fig:mbpthe} and \ref{fig:mbptox} for 
$^{4}$He and $^{16}$O, respectively.

There are several features to be noted. First of all, both figures show that the results
seem to stabilize between seven and eight  
major shells. For $^{4}$He all possible excitations within these shells
are allowed in the computation of the diagrams, to be contrasted to the work of 
Kassis \cite{kassis72} and other traditional approaches \cite{hko95} where one typically
considers only $2-4\hbar\omega$ excitations. For $^{16}$O we keep the same numbers
of maximum allowed excitations for both the $0s_{1/2}$ shell and the $0p_{3/2}0p_{1/2}$ shells.
As an example, 
for eight major shells we could have $14\hbar\omega$ $2p-2h$ excitations from the $0s_{1/2}$  shell
whereas from the   $0p_{3/2}0p_{1/2}$ shell we can at most have $12\hbar\omega$ $2p-2h$
excitations. The latter fixes the total number of allowed excitations with eight major
shells for $^{16}$O. 
The fact that the energies seem to converge at this level of truncation is a welcome
feature which can be exploited in the coupled-cluster calculations. These calculations,
see below, are much more challenging from a computational point of view since we in principle
generate a much larger class of diagrams. In this work we are 
limited to computations
with at most seven major shells in helium and six major shells in 
oxygen in our
coupled-cluster calculations. This 
means that hopefully the trend seen in 
Figs.~\ref{fig:mbpthe} and \ref{fig:mbptox} allows us 
to limit our coupled-cluster
calculations to six or seven major shells. 
Secondly, although the minimum shifts a little 
as function of the oscillator energy as we increase the
oscillator space, we notice
that as the number of major shells is increased, the 
dependence of the binding energy 
upon the oscillator parameter weakens. A similar 
feature is seen in the coupled-cluster calculations
below.  
For $^{16}$O the minimum for seven shells 
takes place at $E/A=-7.155$ MeV for $\hbar\omega = 12.9$ MeV 
and for eight shells we have   $E/A=-7.120$ MeV at $\hbar\omega = 13.6$ MeV. For six shells
we obtain  $E/A=-7.416$ MeV at $\hbar\omega = 12.6$ MeV. 
The curvature for larger values of $\hbar\omega$ decreases with increasing number of shells $N$.
At $\hbar\omega =18$~MeV we have for $^{16}$O and $N=8$ that $d(E/A)/d\omega=0.217$, for 
$N=7$ we obtain $d(E/A)/d\omega=0.277$
and $N=6$ we have $d(E/A)/d\omega=0.346$. 
The corresponding derivatives  for $^{4}$He at $\hbar\omega=18$ MeV
are  for $N=8$ $d(E/A)/d\omega=0.119$, for 
$N=7$ $d(E/A)/d\omega=0.158$
and for $N=6$ $d(E/A)/d\omega=0.209$, indicating a smoother dependence upon $\hbar\omega$ with
increasing $N$.
The reader should also note that 
in the limit $\hbar\omega \rightarrow 0$ we have
$E\rightarrow 0$.  
Finally, we observe that we are not able to 
reproduce the experimental binding energies.
We will come back to this point in Sec.~\ref{sec:results} where a comparison
with coupled cluster theory is also made.

\section{Coupled clusters in single and double excitations}
\label{sec:ccsd}

In this section, we will discuss a formal
derivation of the coupled-cluster equations.
While this discussion is
standard in quantum chemistry (see, for example Ref.~\cite{comp_chem_rev00}), 
the nuclear physics community may
not be familiar with this formulation of coupled-cluster theory. 
We will therefore introduce notations and equations that we 
will continue to use throughout our discussion (in this and 
following papers) of this technique and its extensions.

We must first bring the Hamiltonian into normal ordered
form with respect to the reference state $\mid\Psi_0\rangle$. 
The Hamiltonian then becomes 
\begin{widetext}
\begin{eqnarray}
H_{\rm eff}(\omega) &=& t+G(\omega) =\sum_{pq}f_{pq}\{a_p^\dagger a_q\}
+\frac{1}{4}\sum_{pqrs}\langle pq \mid\mid rs \rangle
\{a_p^\dagger a_q^\dagger a_s a_r\}\nonumber \\ 
&+& \langle\Psi_0\mid H_{\rm eff}(\omega)\mid\Psi_0\rangle=H_N+E_0\;,
\end{eqnarray}
\end{widetext}
where $E_0=\langle\Psi_0\mid H_{\rm eff}(\omega)\mid\Psi_0\rangle$ and
the Fock-matrix element is given by
\begin{equation}
f_{pq}=\langle p\mid t\mid q \rangle +\sum_i \langle pi\mid\mid ri \rangle\;.
\end{equation}
To simplify notation, we will use $\langle pq \mid\mid rs \rangle = 
\langle pq \mid G(\omega)\mid rs \rangle$. 
We employ bracket notation $\{\}$ to indicate normal ordering with respect
to the reference state. As stated previously, we use  the notation that
$p,q,r,s$ refers to  all single-particle states
and $i,j,k,l$ index all sums below the fermi 
surface.  In addition we let 
$a,b,c,d$ index all sums
above the fermi surface.
The total number of single-particle states in the model
space is $N_s = N_p+N_h$ where $N_p$ refers to the number of 
particle states, and $N_h$ is the number of hole states. 
Hereafter we simply use $H$ for $H_{\rm eff}(\omega)$. 

Formally, the coupled-cluster method begins by postulating that 
the correlated many-body wave function is given by 
\begin{equation}
\mid \Psi \rangle = \exp\left(T\right)\mid \Psi_0\rangle  \;,
\end{equation}
where we define the 
the correlation operator as 
\begin{equation}
T=T_1 + T_2 + T_3 + \cdots + T_A \;. 
 \end{equation}
The correlation operators are defined in terms of 
$n$-particle $n$-hole ($n$p-$n$h) excitation amplitudes as 
\begin{eqnarray}
T_1 &=& \sum_{i<\varepsilon_f,a>\varepsilon_f} t^a_i a^+_a a_i\;, \\
T_2 &=& \sum_{i,j<\varepsilon_f; ab > \varepsilon_f}t^{ab}_{ij}
a^+_a a^+_b a_j a_i \;,
\end{eqnarray}
and higher order terms for $T_3$ to $T_A$. 
Coupled-cluster theory may thus be hierarchically  improved upon by 
increasing the number of $T_i$ operators one computes. We will call
the theory in which only $T_1$ and $T_2$ operators are present, CCSD, 
or coupled-clusters at the single and double excitation level. CCSDT 
means that $T_3$ is retained in the correlation operator, while CCSDTQ
refers to keeping both $T_3$ and $T_4$ correlation operators. 
In this uncoupled representation, the correlation amplitudes must 
obey the fermion-symmetry relations which for the 
$T_2$ correlation operators yield $t_{ij}^{ab} = - t_{ji}^{ab} 
= - t_{ij}^{ba} = t_{ji}^{ba}$.  We will use the short-hand notation 
$t_1$ and $t_2$ to represent the 
array of all $1p$-$1h$ and $2p$-$2h$ operators.

We compute the expectation value of the energy from 
\begin{equation}
E=\langle\Psi_0\mid \exp\left(-T\right) H \exp\left(T\right)
\mid\Psi_0\rangle\;. 
\end{equation}
Because the energy is computed using projective, asymmetric techniques,
an important question concerns the physical reality of the coupled-cluster
energy. Quantum mechanics requires that physical observables should be
expectation values of Hermitian operators. The coupled-cluster energy
expression contains the non-Hermitian operator $\left[\exp(-T)H\exp(T)\right]$. 
However, if $T$ is not truncated, the similarity-transformed operator
exhibits an energy-eigenvalue spectrum that is identical to the original
Hermitian operator, $H$, thus justifying its formal use. 
From a practical point of view, the coupled-cluster energy tends to
follow the expectation value result (if the theory is reformulated as 
a variational theory), even when $T$ is truncated. 

The correlation energy is quite easy to calculate and is given by
\begin{widetext}
\begin{equation}
E_{\rm corr}=E-E_0 = \sum_{ia}f_{ia}t^a_i
+\frac{1}{4}\sum_{aibj}\langle ij \mid\mid ab \rangle t^{ab}_{ij}
+\frac{1}{2}\sum_{aibj}\langle ij \mid\mid ab \rangle t^a_i t^b_j \;.
\label{ccsd_energy}
\end{equation}
\end{widetext}
For two-body Hamiltonians, this equation is general and is not 
restricted to the CCSD approximation since higher-order cluster
operators such as $T_3$ and $T_4$ cannot produce fully contracted
terms with the Hamiltonian and therefore contribute zero to the 
energy. Higher-order operators can contribute to the energy 
indirectly through the equations used to determine these 
amplitudes. The three terms in Eq.(\ref{ccsd_energy}) are 
usually referred to as the $T_1$, $T_2$, and $T_1^2$ contributions
to the correlation energy. 

The equations for amplitudes are found by left projection of 
excited Slater determinants
so that 
\begin{eqnarray}
0 &=& \langle\Psi_i^a\mid 
\exp\left(-T\right) H_N \left(T\right) \mid \Psi_0\rangle\;,  \\
0 &=& 
\langle\Psi_{ij}^{ab}\mid 
\exp\left(-T\right) H_N \left(T\right) \mid \Psi_0\rangle \;.
\end{eqnarray}
The Baker-Hausdorf 
relation may be used to rewrite the similarity
transformation as 
\begin{eqnarray}
& & \exp\left(-T\right) H_N \left(T\right)=
H_N+\left[H_N,T_1\right]+\left[H_N,T_2\right] \nonumber \\
& &+\frac{1}{2}\left[\left[H_N,T_1\right],T_1\right] 
+\frac{1}{2}\left[\left[H_N,T_2\right],T_2\right] \nonumber \\
& &+\left[\left[H_N,T_1\right],T_2\right]+\cdots \;.
\end{eqnarray}
The expansion terminates exactly at quadruply nested commutators  when
the Hamiltonian contains at most two-body terms, and at six
nested commutators when  three-body terms are present. We stress that
this termination is exact, thus allowing for a derivation of exact
expressions for the $t_1$ and $t_2$ amplitudes. To derive these equations
is straightforward but tedious work.  For example, one of the amplitude
terms is given by
\begin{eqnarray}
 & & \langle\Psi^{ab}_{ij}\mid \left(G_N T_1^2 T_2 \right)_c\mid \Psi_0\rangle
=\sum_{pqrs}\sum_{kc}\sum_{ld}\sum_{mnef} 
t^c_k t^d_l t^{ef}_{mn} \nonumber \\
& &
\langle\Psi_0\mid \{a^+_i a^+_j a_b a_a\}\{a^+_p a^+_q a_s a_r\} \nonumber \\
& & \{a^+_c a_k\}
\{a^+_c a_k\}\{a^+_d a_l\}\{a^+_e a^+_f a_n a_m\}\mid\Psi_0\rangle  \;,
\end{eqnarray}
for which Wick's theorem may be used to calculate the expectation 
matrix element.  Here $G_N$ is the normal ordered $G$-matrix operator, and the 
subscript $c$ indicates only connected diagrams enter into the computation
of the expectation value. 

The $t_1$ amplitude equations are given by
\begin{widetext}
\begin{eqnarray}
0 & = & f_{ai} + \sum_cf_{ac}t^c_i - \sum_k f_{ki}t^a_k +  
\sum_{kc}\langle ka\mid\mid ci\rangle t^c_k + \sum_{kc}f_{kc}t^{ac}_{ik}
+\frac{1}{2}\sum_{kcd}\langle ka\mid\mid cd\rangle t^{cd}_{ki} \nonumber \\
  & - & \frac{1}{2}\sum_{klc}\langle kl\mid\mid ci \rangle t^{ca}_{kl}
-\sum_{kc}f_{kc}t^c_it^a_k -\sum_{klc}\langle kl\mid\mid ci\rangle t^c_k t^a_l
+\sum_{kcd} \langle ka\mid\mid cd \rangle t^c_k t^d_i  \nonumber \\
  & - & \sum_{klcd}\langle kl\mid\mid cd\rangle t^c_kt^d_it^a_l
+ \sum_{klcd}\langle kl\mid\mid cd\rangle t^c_kt^{da}_{li}
- \frac{1}{2} \sum_{klcd}\langle kl\mid\mid cd\rangle t^{cd}_{ki}t^{a}_{l}
- \frac{1}{2} \sum_{klcd}\langle kl\mid\mid cd\rangle t^{ca}_{kl}t^{d}_{i}\;.
\label{t1_eqn}
\end{eqnarray}
\end{widetext}
This equation is non-linear in the $t_1$ amplitudes, and linear in
the $t_2$ amplitudes. 

The $t_2$ amplitude equations are given by 
\begin{widetext}
\begin{eqnarray}
0 & = & \langle ab \mid\mid ij \rangle 
+ \sum_c\left(f_{bc}t^{ac}_{ij} - f_{ac}t^{bc}_{ij}\right)
- \sum_k\left(f_{kj}t^{ab}_{ik} - f_{ki}t^{ab}_{jk}\right)  \nonumber \\
  & + & \frac{1}{2}\sum_{kl}\langle kl \mid \mid ij\rangle t_{kl}^{ab}
+ \frac{1}{2}\sum_{cd}\langle ab \mid \mid cd\rangle t_{cd}^{ij}
+ P(ij)P(ab)\sum_{kc}\langle kb \mid \mid cj\rangle t_{ac}^{ik} \nonumber \\
  & + & P(ij)\sum_c \langle ab \mid\mid cj\rangle t^c_i
- P(ab)\sum_k \langle kb \mid\mid ij\rangle t^a_k \nonumber \\
  & + & \frac{1}{2}P(ij)P(ab)\sum_{klcd} \langle kl\mid\mid cd \rangle 
        t_{ik}^{ac}t_{lj}^{db} 
+ \frac{1}{4}\sum_{klcd} \langle kl\mid\mid cd \rangle 
        t_{ij}^{cd}t_{kl}^{ab} 
- \frac{1}{2}P(ab)\sum_{klcd} \langle kl\mid\mid cd \rangle 
        t_{ij}^{ac}t_{kl}^{bd} 
- \frac{1}{2}P(ij)\sum_{klcd} \langle kl\mid\mid cd \rangle 
        t_{ik}^{ab}t_{jl}^{cd}  \nonumber \\
  & + & \frac{1}{2}P(ab)\sum_{kl}\langle kl \mid\mid ij\rangle t^a_kt^b_l
+ \frac{1}{2}P(ij)\sum_{cd}\langle ab\mid\mid cd\rangle t^c_i t^d_j
- P(ij)P(ab)\sum_{kc}\langle kb\mid\mid ic\rangle t^a_k t^c_j \nonumber \\
  & + & P(ab) \sum_{kc} f_{kc} t^a_k t^{ab}_{ij} 
+  P(ij) \sum_{kc} f_{kc} t^c_i t^{ab}_{jk}  \nonumber \\
  & - & P(ij)\sum_{klc} \langle kl\mid\mid ci\rangle t^c_k t^{ab}_{lj}
+  P(ab)\sum_{kcd} \langle ka\mid\mid cd\rangle t^c_k t^{db}_{ij}
+  P(ij)P(ab)\sum_{kcd} \langle ak\mid\mid dc\rangle t^d_i t^{bc}_{jk}
   \nonumber \\
  & + & P(ij)P(ab)\sum_{klc} \langle kl\mid\mid ic\rangle t^a_l t^{bc}_{jk}
+ \frac{1}{2} P(ij)\sum_{klc} \langle kl\mid\mid ck\rangle t^c_i t^{ab}_{kl}
- \frac{1}{2} P(ab)\sum_{kcd} \langle kb\mid\mid cd\rangle t^a_k t^{cd}_{ij}
  \nonumber \\
  & - & \frac{1}{2} P(ij)P(ab)\sum_{kcd}\langle kb\mid\mid cd \rangle 
         t^c_it^a_kt^d_j
+ \frac{1}{2} P(ij)P(ab)\sum_{klc}\langle kl\mid\mid cj \rangle 
         t^c_it^a_kt^b_l \nonumber \\
  & - & P(ij)\sum_{klcd} \langle kl \mid\mid cd \rangle t^c_k t^d_i t^{ab}_{lj}
- P(ab)\sum_{klcd} \langle kl \mid\mid cd \rangle t^c_k t^a_l t^{db}_{ij}
+ \frac{1}{4}P(ij)\sum_{klcd} \langle kl \mid\mid cd \rangle 
    t^c_i t^d_j t^{ab}_{kl} \nonumber \\
  & + & \frac{1}{4}P(ab)\sum_{klcd} \langle kl \mid\mid cd \rangle 
    t^a_k t^b_l t^{cd}_{ij} 
+ P(ij)P(ab)\sum_{klcd} \langle kl \mid\mid cd \rangle 
t^c_i t^b_l t^{ad}_{kj}
+ \frac{1}{4}P(ij)P(ab)\sum_{klcd} \langle kl \mid\mid cd \rangle 
  t^c_i t^a_k t^d_j t^b_l \;.
\label{t2_eqn}
\end{eqnarray}
\end{widetext}
The permutation operator $P$ yields
\begin{equation}
P(ij)f(ij) = f(ij) - f(ji)
\end{equation}
The equations of the $t_2$ amplitudes are nonlinear in both $t_1$ and
$t_2$ terms. While these equations appear quite lengthy, they are 
solvable through iterative techniques that we will discuss below. 
We note that the amplitude equations include terms that allow for 
4p-4h excitations. Indeed, while we speak of doubles in terms of 
amplitudes, the class of diagrams involved in the theory include 
fourth-order terms. This is a very important difference 
and distinction between the shell model with up to 
$2p$-$2h$ excitations and CCSD. Furthermore, when
the energy is computed in CCSD, all terms are linked and connected. 

In order to calculate expectation values of operators we use the 
Hellmann-Feynman theorem \cite{feynman} which states that if we 
perturb our Hamiltonian such that $H'=H+\lambda \Omega$ where $\lambda$ is a
small quantity and $\Omega$ is the operator of interest, then the
energy changes only by a small amount from its original value of 
$E(\lambda=0)$.  As a function of $\lambda$, the energy becomes
$E'=E(\lambda=0)+\lambda dE/d\lambda$, 
and the expectation value of the operator is given by 
\begin{equation}
\langle \Omega \rangle =\frac{dE(\lambda=0)}{d\lambda} \;.
\label{helfeyn}
\end{equation}

\section{CCSD calculations for helium and oxygen}
\label{sec:results}

\subsection{Iteration of the equations}

Several computational challenges arise when we implement the CCSD
equation solver. One problem involves memory requirements for the 
$G$-matrix, which in its uncoupled form is a 4-index tensor and 
therefore requires a large amount of storage. In order to maintain
fast computation, we do not employ storage compression of the $G$-matrix
at this stage. Thus, for example, 
an $N=7$ calculation requires 100 GBytes of storage for
the $G$-matrix elements. 
Present-day parallel 
computing systems use distributed memory architectures that allow for
the storage of such large sets of data. Our implementation in distributing
the $G$-matrix is to store the third and fourth indices across processors in
sub-matrix blocks. In doing this, we are able to take advantage of 
the large available memory for storing the matrix elements. 

We use an iterative method to generate 
solutions to the $t_1$ and $t_2$ amplitude equations. In this paper,
we are concerned with closed shell nuclei. In this case, a slight 
rearrangement of Eqs.~(\ref{t1_eqn}) and (\ref{t2_eqn}) gives a second
order perturbative initial solution for the amplitudes. For the 
$t_1$ amplitude, we rearrange the first few terms by pulling out the
diagonal in the first two sums and defining a $1p$-$1h$ and $2p$-$2h$ 
energy denominators as
\begin{eqnarray}
D^a_i & = & f_{ii}-f_{aa} \;, \\
D^{ab}_{ij} & = & f_{ii}+f_{jj}-f_{aa}-f_{bb} \;.
\end{eqnarray}
The first terms in the $t_1$ and $t_2$ amplitude equations then become
\begin{widetext}
\begin{eqnarray}
D^a_i t^a_i & = & f_{ai} + 
+ \sum_c (1-\delta_{ca})f_{ac}t^c_i - \sum_k (1-\delta_{ik})f_{ik}t^a_k 
+\cdots \;, \\
D^{ab}_{ij}t^{ab}_{ij} & = & \langle ab\mid\mid ij \rangle +
P(ab)\sum_c(1-\delta_{bc})f_{bc}t^{ac}_{ij}+
P(ij)\sum_k(1-\delta_{kj}) f_{kj}t^{ab}_{ij}+\cdots\;.
\end{eqnarray}
\end{widetext}
While these are exactly the same equations as given in 
Eqs.~(\ref{t1_eqn}) and (\ref{t2_eqn}),
we may use them to begin an iterative solution by 
initially setting all amplitudes on the left-hand side to zero. We
then obtain for initial amplitudes 
\begin{eqnarray}
t^a_i &=& \frac{f_{ai}}{D^a_i} \\
t^{ab}_{ij} &=& \frac{\langle ab \mid\mid ij \rangle}{D^{ab}_{ij}}\;. 
\label{amp_start}
\end{eqnarray}

We now need to compute the various terms in Eq.~(\ref{t2_eqn}) to obtain
the new amplitudes. Since our $G$-matrix elements are distributed across
processors, we will have partial sums on certain indices. For clarity,
we do not show this complication in the following. 
We demonstrate the numerical procedure by considering one of the terms
in the two-particle-two-hole amplitude equation, Eq.~(\ref{t2_eqn}):
\begin{equation}
f(ab,ij)= \sum_{kl,cd} \langle kl \mid \mid cd \rangle t^{cd}_{ij}t^{ab}_{kl}
\end{equation}
We perform the following mapping to matrices $M$, $N$, and $O$, $P$, and $Q$:
\begin{eqnarray}
M_{\alpha,\beta} &=& t^{ab}_{kl} \nonumber \\
N_{\beta,\gamma} &=& \langle k l \mid\mid cd \rangle \nonumber \\
O_{\gamma,\delta}&=& t^{cd}_{ij}  \nonumber \\
Q_{\alpha,\delta}&=& f(ab,ij)
\end{eqnarray}
where we map the indices $\alpha=(a,b)$, $\beta=(k,l)$, $\gamma=(c,d)$, and
$\delta=(i,j)$. Our computation then becomes two matrix-matrix multiplications:
\begin{eqnarray}
P_{\alpha,\gamma}=\sum_{\beta}  M_{\alpha,\beta} N_{\beta,\gamma} \nonumber \\
Q_{\alpha,\delta}=\sum_{\gamma} P_{\alpha,\gamma}O_{\gamma,\delta}
\end{eqnarray}
followed by a mapping of $Q_{\alpha,\delta}$ to $f(ab,ij)$. 
All terms in the amplitude equations can be formulated
in this way and three important consequences follow. 
First, the computational work 
scales from one computation of order ${\cal O}(N_p^4 N_h^4)$ to two computations
of ${\cal O}(N_p^4 N_h^2)$. Second, by formulating the algorithm in this 
manner, we take advantage of highly optimized Basic Linear Algebra 
Subprograms (BLAS) \cite{blas}.  
This represents a significant reduction in effort when
contrasted with the naive equations. We see that a calculation $^{16}$O 
should require sixteen times more computational time when compared to a
calculation of $^{4}$He in the same model space. Due to the longer loop
structure in $^{16}$O the time required to complete a single oxygen run is
only a factor of ten larger than the He case. 
Third, because we have sub-blocked the interaction amplitudes, we 
can spread the work across numerous processors. In doing so, we must 
perform a global reduction operation (a global sum) in order to 
generate the new amplitudes for the next iteration. 
We show in Table \ref{tab_computational}
some details of the computational sizes of the $m$-scheme problems 
that we have undertaken in this paper. The number of unknowns for
which we are solving is approximately the number of two-particle-two-hole
amplitudes. Therefore, in our largest calculation for $^{16}$O we 
are actually solving approximately 176,300 equations per iteration. 
We also list the number of nonzero $G$-matrix elements in Table 
\ref{tab_computational}.

\begin{table}
\begin{center}
\begin{tabular}{|ccccc|}
\hline
N & $N_s$ & $^4$He & G (in millions) & $^{16}$O \cr
\hline
4 & 80  &  1,792  & 0.93 & 24,960 \cr
5 & 140 &  4,000  & 7.23 & 77,880 \cr
6 & 224 &  7,976  & 40.4 & 176,240 \cr
7 & 336 & 14,112  & 178.5 &  --    \cr
\hline
\end{tabular}
\end{center}
\caption{Various memory scalings for the He and O systems as
a function of the number of major oscillator shells, $N$. Listed
are the number of uncoupled single-particle states, the two-particle-two-hole
amplitudes in $^{4}$He, the number of non-zero $G$-matrix elements, and
the number of nonzero two-particle-two-hole amplitudes in $^{16}$O. }
\label{tab_computational}
\end{table}

Eq.~(\ref{amp_start}) represents the terms
that one uses to compute the energy
in second-order perturbation theory of the 
M{\o}ller-Plesset type \cite{mp34}. We show in Table~\ref{e_tab}
the energy obtained from the $0^{th}$ $E(0)$ iteration and the final
iteration $E(F)$ as a function of increasing oscillator levels in the $^{16}$O
system. Note that the difference between the converged CCSD energies
and the initial $0^{th}$ order energies increases as the basis
space increases. The converged summation of the CCSD 
equations yields approximately
10 MeV (or 0.6~MeV per particle) in extra binding. 
These findings are corroborated by those from many-body perturbation theory from Sec.~\ref{sec:mbpt}. 
It is therefore worth comparing 
these results with those from second-order and third-order 
many-body perturbation theory as well. These are labeled 
$E^{\mathrm{2nd}}_{\mathrm{MBPT}}$ and $E^{\mathrm{3rd}}_{\mathrm{MBPT}}$
in the same table. The reader should notice that the zeroth iterations of the 
coupled-cluster schemes already includes corrections to the one-body amplitudes
$t_1$. However, the energy 
denominators used in the computation of the second-order
diagrams of Fig.~\ref{fig:diagrams} (diagrams 2 and 3) 
have hole states determined by Eq.~(\ref{eq:poormansbhf}). The agreement with the zeroth order iteration
and second-order perturbation theory is very good, especially for 
five and six major shells, as can be seen from Table \ref{e_tab}.
\begin{table}
\begin{center}
\begin{tabular}[t]{|ccccc|}
\hline
N ($\hbar\omega)$ & $E(0)$  & $E^{\mathrm{2nd}}_{\mathrm{MBPT}}$& $E^{\mathrm{3rd}}_{\mathrm{MBPT}}$&$E(F)$ \cr
\hline
4 14  & -135.117  &  -132.063 & -129.920 &-140.473 \cr
5 14  & -124.786  &  -124.844 & -121.520 & -127.790 \cr
6 14   & -121.356  &  -121.481& -118.233 & -119.733 \cr
\hline
\end{tabular}
\end{center}
\caption{Comparisons of the 0$^{th}$ order energy $E(0)$ 
and the converged CCSD results $E(F)$ for $\beta_{\rm c.m}=0.0$ 
in $^{16}$O as a function of increasing model model space.
The results are also compared with many-body perturbation theory to second and third order, $E^{\mathrm{2nd}}_{\mathrm{MBPT}}$ and $E^{\mathrm{3rd}}_{\mathrm{MBPT}}$, respectively. All energies are in MeV.}
\label{e_tab}
\end{table}
However, for third-order perturbation theory one clearly sees fairly large 
differences compared with the coupled-cluster results.
Typically, the relation between first- and second-order in perturbation theory
for  $^{16}$O is given by a factor of $\sim 5-6$. For e.g., $N=5$ and 
$\hbar\omega=14$ MeV, we have $-329.124$ MeV from first order and $-47.719$
from second order. To third order we obtain a repulsive contribution of
$3.324$ MeV, to be contrasted with the almost $3$ MeV of attraction given 
by higher-order terms in the coupled-cluster expansion. This indicates that
many-body perturbation theory to third order is
most likely not a converged result. An interesting feature to be noted from
many-body perturbation theory calculations is that higher terms loose their
importance as the size of system is increased. For 
$^{4}$He the relation between first-order and second-order perturbation theory
is given by a factor of $\sim 3-4$, depending on the value of $\hbar\omega$.
Calculations for $^{40}$Ca not reported here indicate a relation of
$\sim 7-9$ between first-order and second-order perturbation theory.
This is somewhat expected since the $G$-matrix is smaller for larger
systems, although the energy denominators becomes smaller.  

\begin{figure}
\vspace{0.25in}
\includegraphics[angle=270, scale=0.35]{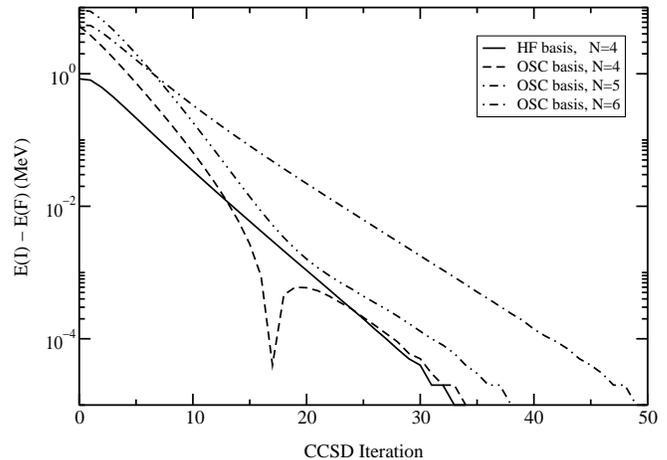}
\caption{The convergence of the ground-state energy
as a function of the CCSD iterations for $^{16}$O.}
\label{converge_fig}
\end{figure}

We show typical convergence curves for our coupled-cluster calculations 
in Fig.~\ref{converge_fig} for $^{16}$O for the $N=4,5,6$ oscillator
shell model spaces. In this figure, $E(I)$ is the energy at the $I^{th}$
iteration, while $E(F)$ is the final energy of the system. We obtain
convergence within 50 iterations in most cases. The curves also 
exhibit an exponentially convergent behavior. 

Since the CCSD amplitude equations are nonlinear, we investigated
the results obtained by using different initial reference
states. The naive choice for $\mid\Psi_0\rangle$ is to fill the 
lowest oscillator states for a given system. For example, we define 
$\mid\Psi_0\rangle$ as the filled $0s$ state for $^{4}$He and the
filled $0s$-$0p$ states for $^{16}$O. This choice is fine for closed
shell nuclei since none of the energy denominators discussed above becomes
zero, but for other nuclei this would be a problem. As an alternative 
procedure, we start with the Hartree-Fock ground-state Slater determinant
for a given system. We solve the Hartree-Fock equations in the oscillator
basis in order to obtain transformation matrices $D$ that take us from the
oscillator to the Hartree-Fock basis. We then transform the Hamiltonian 
to the Hartree-Fock basis using the relation 
$a_i^\dagger = \sum_{\alpha}D_{\alpha i}c_\alpha^\dagger$ where $a_i$ and
$a_i^\dagger$ annihilate and create particles in the oscillator basis and
$c_\alpha$ and $c_\alpha^\dagger$ annihilate and create particles in the 
Hartree-Fock basis. Note that $D^\dagger D =1$. While a complete 
diagonalization of $H$ in either basis would yield the same results, 
the CCSD amplitude equations are not invariant under this 
transformation since states below and above the fermi surface will be 
mixed. Furthermore, at the Hartree-Fock level, rotational symmetry of 
the Hamiltonian is broken, although correlations including those
at the CCSD level
will restore much of this symmetry. Although we do not discuss in this 
paper open-shell systems, the Hartree-Fock solution offers a clean way
obtain a reference Slater determinant for those cases. 

We show in Fig.~\ref{converge_fig} the convergence of the 
CCSD equations in both the oscillator and Hartree-Fock basis for 
$N=4$ major shells. In both cases, convergence at the $10^{-5}$ level
is reached before 40 iterations. A more qualitative assessment 
of how the choice of reference state affects the final results
is seen in Table~\ref{table1}. In the oscillator basis, the $1p-1h$ amplitudes
carry a significant fraction of the correlation energy, while in the 
Hartree-Fock basis these terms do not contribute to the energy.  The difference
between the Hartree-Fock and oscillator total energies is 0.6\%. 
We note that if one
completely ignores the $1p-1h$ amplitudes in the Hartree-Fock basis, the
energy becomes -138.38 MeV, a difference of less than 0.07\%. In the
oscillator basis, a much larger error of 5.6\% is obtained if one ignores
the $1p-1h$ amplitudes. Furthermore, the CCD equations (ignoring the 
$1p-1h$ amplitudes) sometimes exhibit numerical instabilities similar to
ignoring the Hartree-Fock insertions 
in many-body perturbation theory discussed above. 

\begin{table}
\begin{center}
\begin{tabular}{|ccc|}
\hline
Term & Oscillator (MeV) & Hartree-Fock (MeV) \cr
\hline
$E_0$   & -109.45 & -121.977 \cr
$T_1$   & -9.669  & $7\times 10^{-6}$ \cr
$T_1^2$ & -1.757  & $-0.3\times 10^{-3}$ \cr
$T_2$   & -18.439 & -16.495  \cr
$E_{\rm corr}$  & -29.865  & -16.498 \cr
$E_{\rm Total}$ & -139.31  & -138.47 \cr
\hline
\end{tabular}
\end{center}
\caption{Comparisons of CCSD results in $^{16}$O when using naively
filled oscillator reference state, or when using the Hartree-Fock 
reference state.These calculations were performed at $\hbar\omega=14$~Mev
and $\beta_{\rm c.m.}=1.0$. }
\label{table1}
\end{table}

\subsection{$^{4}$He and $^{16}$O ground states}

\begin{figure}
\vspace{0.25in}
\includegraphics[angle=270, scale=0.35]{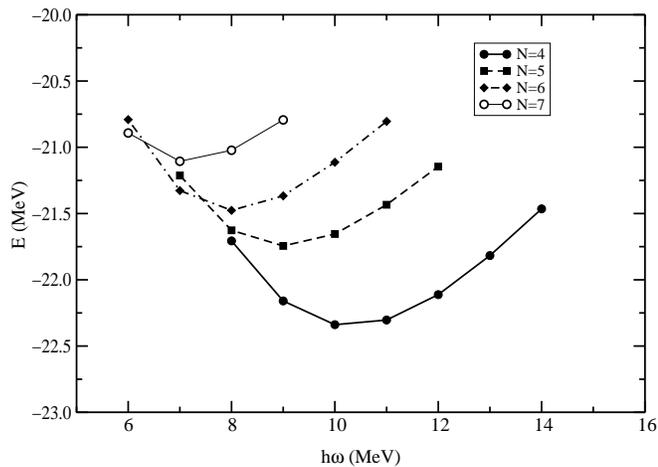}
\caption{Dependence of the ground-state energy of $^{4}$He  on $\hbar\omega$
as a function of increasing model space.}
\label{fig_he_hw}
\end{figure}

We now return to a discussion of $^{4}$He and $^{16}$O by providing
a description of their ground-state energies using the CCSD formalism.
As in the many-body perturbation theory section of this paper, we wish to demonstrate
how the coupled-cluster theory converges as a function of increasing model 
space. We are currently able to perform 
this study for up to seven major oscillator shells in helium and up to 
six shells in oxygen. 
While these studies do not address the starting-energy
dependence of the $G$ matrix, they do indicate the convergence of the 
calculations as a function of model space, and they indicate the softening
of the $\hbar\omega$ dependence as one moves to larger spaces. 
The $\hbar\omega$ 
dependence of the ground state energy of $^{4}$He 
as a function of the model space is
shown in Fig.~\ref{fig_he_hw} for the $N=4,5,6,7$ major oscillator shells with 
$\beta_{\rm c.m.}=0$. These curves generally exhibit a parabolic character. 

\begin{figure}
\vspace{0.25in}
\includegraphics[angle=270, scale=0.35]{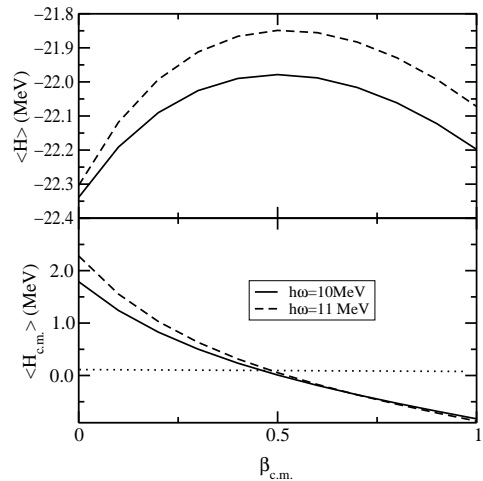}
\caption{Top panel: the total energy of $^4$He in four major 
oscillator shells plotted as a function of the center-of-mass 
parameter $\beta_{\rm c.m.}$. Bottom panel: the center of mass energy
as a function of $\beta_{\rm c.m.}$. The energy is at a variational 
minimum in this 4 oscillator shell space at $\hbar\omega=10$~MeV. 
The variational minimum for
as a function of the CCSD iterations.} 
\label{fig_com}
\end{figure}

We applied the center-of-mass correction described above to both the
He and O nuclei. We demonstrate 
how this procedure behaves when one solves the CCSD equations in
Fig.~\ref{fig_com} for $^{4}$He in the $N=4$ and model space. 
We calculated the expectation value of 
of $H_{\rm c.m.}$ with Eq.~(\ref{helfeyn}). These plots indicate
that the center-of-mass behavior is less severe when we perform 
calculations at the variational minimum in $\hbar\omega$, which in this
case is at $\hbar\omega=10$~MeV. We observe that these corrections 
are rather small for both He and O, amounting to a correction of less
than 1\% in both cases. This correction only applies to the ground
state. 

We performed a full diagonalization of $^4$He in four major oscillator
shells at $\hbar\omega=10$~MeV and $\beta_{\rm c.m.}=0.5$
and obtained an energy of $-23.4837$~MeV for the ground state as compared
to our CCSD result of $-21.98$~MeV, a 6.8\% difference. We anticipate that
triples corrections to the energy \cite{ccsdt03} will recover much of the
remaining difference between the complete result and the CCSD result. 
The CCSD results should be approximately of similar quality in larger
model spaces so that we can expect an approximate complete diagonalization
energy of $-22.4$~MeV in seven oscillator shells as compared to 
$-21.0$~MeV for the CCSD result. The ground-state energy using
Idaho-A was quoted as -27.4~MeV by Navratil and Ormand in
Ref.~\cite{petr_erich02}. The
major difference between our projected full diagonalization result 
and -27.4~MeV is most likely due to the differing 
treatments of the $G$-matrix. 

\begin{figure}
\vspace{0.25in}
\includegraphics[angle=270, scale=0.35]{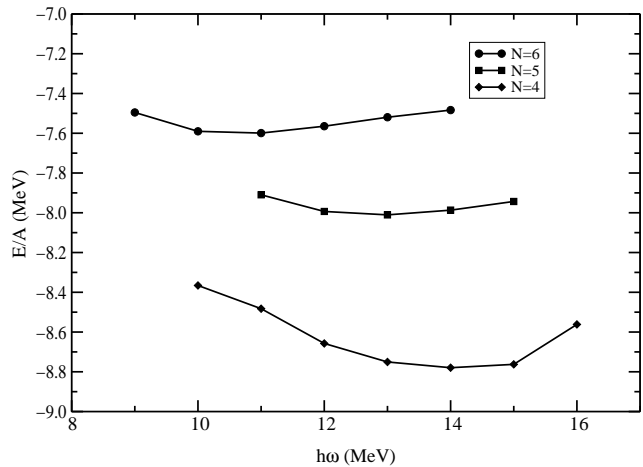}
\caption{Dependence of the ground-state energy of $^{16}$O  on $\hbar\omega$
as a function of increasing model space.}
\label{fig_ox_hw}
\end{figure}

In this initial study we performed calculations of the $^{16}$O 
ground state for up to six major oscillator shells as a function
of $\hbar\omega$. Fig.~\ref{fig_ox_hw} indicates the level of convergence
of the energy per particle for $N=4,5,6$ shells. The experimental value
resides at 7.98~MeV per particle. While this calculation is not completely
converged, it does show signs of convergence as the space is increased. 
By six oscillator shells, the $\hbar\omega$ dependence becomes rather
minimal. We find a ground-state binding energy of 7.6 MeV per particle in
oxygen using the Idaho-A potential. Since the Coulomb interaction should give
approximately 1 MeV/A of repulsion, and is not included in this 
calculation, we actually obtain approximately 6.6 MeV of nuclear binding
in the 6 major shell calculation which is somewhat above the experimental
value. We note that the entire procedure ($G$-matrix plus CCSD) tends to 
approach from below converged solutions. 

\section{Conclusions and perspectives}
\label{sec:conclusion}

Our goal in this paper has been to describe a nonperturbative method of 
solution to the many-body problem that sums classes of diagrams built upon
low order (up to fourth order in this paper) many-body perturbation theory
diagrams. We have shown how to calculate nuclear ground states using 
coupled-cluster methods. In this paper, 
we concentrated on CCSD equations. 
We used a $G$-matrix as our two-body interaction and 
we expanded in the spherical harmonic-oscillator basis. We reiterate that
CCSD is a nonperturbative approach to the many-body problem: we sum 
classes of diagrams to infinite order in order to calculate various
nuclear properties.  The coupled-cluster method discussed here 
clearly demonstrates the need of summing many-body correlations to infinite order. This is seen when comparing the coupled-cluster results to many-body 
perturbation theory. Furthermore, the latter
is hard to extend beyond third order  
without a self-consistent single-particle potential and beyond 
fourth order with a self-consistent potential.

Before closing
this paper, we would like to discuss several steps that we will take 
during the course of this research. 

One improvement upon the method will be to include the calculation of 
triples excitations (called CCSDT or approximations to it). We indicated 
that the three-particle-three-hole diagrams likely give repulsion and are
important for the description of ground-state properties. Since the CCSD
equations do not include the 3p-3h diagrams completely, and since we have
seen that these diagrams are important, we will eventually need to 
include the triples amplitudes into our equations. Various methods
that include triples diagrams have been investigated by quantum 
chemists and we will investigate which of these methods are appropriate
for the nuclear problem. 

The effort to perform a complete solution to 
the quantum many-body problem grows exponentially 
as one adds particles. Combined with the difficulty
of methods we suffer in nuclear physics from interactions that
are not completely determined. Thus, our methods and techniques
for solution typically develop in lock step with our understanding of
the nuclear Hamiltonian. While several sets of two-nucleon 
of interactions that fit nucleon scattering perfectly have been 
developed over the last 10 years, their many-body characteristics
(in particular, their ability to obtain nuclear ground-state masses) 
indicate that they are insufficient. Three-nucleon interactions become
necessary even to fit the triton and $^{4}$He. To date, no derivations of
CCSD (or CCSDT) equations exist that incorporate a three-body interaction. 
We will pursue this effort in future research. 

CCSD and its extensions can be used to obtain excited state information
by diagonalizing $\bar{H}=\exp\left(-T\right)H\exp\left(T\right)$ in
the space of all singly- and doubly-excited determinants where 
the amplitudes are obtained directly from the converged CCSD
amplitudes.  This will be an important step in the development of 
the coupled-cluster method for nuclear science. 

Finally, our results do depend on the starting energy 
of the effective interaction.
For $^{16}$O and four oscillator shells and $\hbar\omega=14$~MeV, 
our $\beta_{\rm c.m.}=0.0$ result is 
$-140.47$~MeV with the starting energy of $-80$~MeV, while
we obtain $-143.53$~MeV with a starting energy of $-60$~MeV. 
The dependence is therefore weak, but still 
present.  The dependence is more crucial in helium since the binding
energy is much lower, and our $G$-matrix is not defined for positive
starting energies.  There are two possible ways to 
overcome this problem. One is to use the similarity transformation
of Lee and Suzuki, following closely the no-core approach of Barrett, 
Navratil and co-workers \cite{bruce1,bruce2,bruce3}.
This yields a hermitian and starting energy independent interaction 
for a large space. Alternatively, one can compute
self-consistently the single-particle energies using a $G$-matrix 
defined for both positive and negative starting energies. 
Thereafter, a starting energy independent
interaction can be obtained using e.g., the prescription of 
Eq.~(\ref{eq:gmat1approx}), but for both holes
and particles. These approaches will be investigated in future works. 

\section*{Acknowledgments}
Research sponsored by the Laboratory Directed Research and Development
Program of Oak Ridge National Laboratory (ORNL), managed by
UT-Battelle, LLC for the U. S.  Department of Energy under
Contract No. DE-AC05-00OR22725 and the Research Council of Norway. 
We are pleased to acknowledge useful discussions with David Bernholdt,
Ray Bishop, Karol Kowalski, Thomas Papenbrock, Piotr Piecuch, 
and Neils Walet.

\bibliography{ccsd1}

\end{document}